\documentclass[prl,twocolumn,groupedaddress,10pt]{revtex4-1} 
\usepackage{amsmath,amssymb,graphicx}

\usepackage{color,times}

\definecolor{darkblue}{rgb}{0, 0, 0.8}
\usepackage[colorlinks=true, breaklinks=true, linkcolor=darkblue, citecolor=darkblue, urlcolor=darkblue]{hyperref}
\newcommand{\ket}[1]{\left| #1\right\rangle}

\begin{document}

\title{Local optical control of the resonant dipole-dipole interaction between Rydberg atoms}

\author{S. de L\'es\'eleuc$^{\ast}$, D. Barredo$^{\ast}$, V. Lienhard$^{\ast}$, A. Browaeys and T. Lahaye}

\affiliation{Laboratoire Charles Fabry, UMR 8501, Institut d'Optique, CNRS, Univ Paris Sud 11,\\
2 avenue Augustin Fresnel, 91127 Palaiseau cedex, France }

\begin{abstract}

We report on the local control of the transition frequency of a spin-$1/2$ encoded in two Rydberg levels of an individual atom by applying a state-selective light shift using an addressing beam. With this tool, we first study the spectrum of an elementary system of two spins, tuning it from a non-resonant to a resonant regime, where ``bright'' (superradiant) and ``dark'' (subradiant) states emerge. We observe the collective enhancement of the microwave coupling to the bright state. We then show that after preparing an initial single spin excitation and letting it hop due to the spin-exchange interaction, we can freeze the dynamics at will with the addressing laser, while preserving the coherence of the system. In the context of quantum simulation, this scheme opens exciting prospects for engineering inhomogeneous XY spin Hamiltonians or preparing spin-imbalanced initial states.   

\end{abstract}

\maketitle

Real-world magnetic materials are often modeled with simple spin Hamiltonians exhibiting the key properties under study. Despite their simplified character, these models remain challenging to solve, and an actively explored approach is to implement them in pristine experimental platforms~\cite{Georgescu2014}. Such \emph{quantum simulators} usually require an ordered assembly of interacting spins, also called qubits in the case of spin-$1/2$, manipulated by global and local coherent operations. Local operations are a crucial element of a quantum simulator and they have been used, for example, to perform one-qubit rotations for quantum state tomography~\cite{Blatt2008}, to engineer two-qubit quantum gates (see e.g.~\cite{Veldhorst2015,Salathe2015}), or to prepare peculiar initial states~\cite{Choi2016,Marcuzzi2017} and apply local noise~\cite{Smith2015} for studies of many-body localization. To achieve a local operation, one usually shifts the frequency of one targeted qubit in the system. Depending on the physical platform, different approaches are used to accomplish this, such as applying static electric fields for quantum dots~\cite{Veldhorst2014}, or magnetic fluxes for superconducting circuits~\cite{Houck2012}. In atomic systems, focusing an off-resonant laser beam  on a single site can be used to apply an AC-Stark shift on ground-state levels~\cite{Weitenberg2011,Labuhn2014,Xia2015,Wang2015}.

Another promising approach for quantum information science and quantum simulation of spin Hamiltonians are atomic platforms based on Rydberg states~\cite{Saffman2010, Weimer2010}, as they provide strong, tunable dipole-dipole interactions~\cite{Browaeys2016, Zeiher2016, Biedermann2016}. In addition, arrays of optical tweezers allow the efficient preparation of assemblies of up to 50 atoms, arranged in arbitrary geometries, as has been recently demonstrated~\cite{Barredo2016,Endres2016}. One can encode a spin-$1/2$ between the ground-state and a Rydberg level, use the van der Waals interactions between two identical Rydberg states and map the system onto an Ising-like Hamiltonian~\cite{Labuhn2016}. In this case, the spins can be manipulated globally by a resonant laser field and local addressing has been demonstrated using a far red-detuned focused laser beam shifting the ground-state energy of a particular atom in the ensemble~\cite{Labuhn2014}.

In addition to Ising Hamiltonian, the long-range XY Hamiltonian~\cite{Deng2005,Hauke2010,Varney2011,Peter2012,Yan2013,Dalmonte2015} can naturally be implemented with Rydberg atoms by using the dipolar spin-exchange interaction~\cite{Pillet1998,Gallagher1998,Barredo2014,Maxwell2014,Orioli2017}. For principal quantum number $n \sim 60$, the direct dipole-dipole coupling $U = C_3/R^3$  between two atoms in Rydberg levels with orbital angular momentum differing by $\pm 1$ ensures strong interaction energies in the $1-10$~MHz range for atoms separated by $\sim 10 \, \mu$m. The spin-$1/2$ is in this case encoded in two Rydberg levels with a lifetime of a few $100\,\mu$s and a transition frequency in the microwave domain. Due to the exaggerated electric dipole of Rydberg states, high Rabi frequencies are obtained with low microwave power, resulting in a fast coherent manipulation of single atoms. However, for local tuning of the frequency, the previous schemes, shifting a ground-state level~\cite{Labuhn2014}, are irrelevant, and the implementation of a selective Rydberg level shift was so far missing in the toolbox of quantum simulation of XY Hamiltonians.

\begin{figure}
	\centering
	\includegraphics[width=\linewidth]{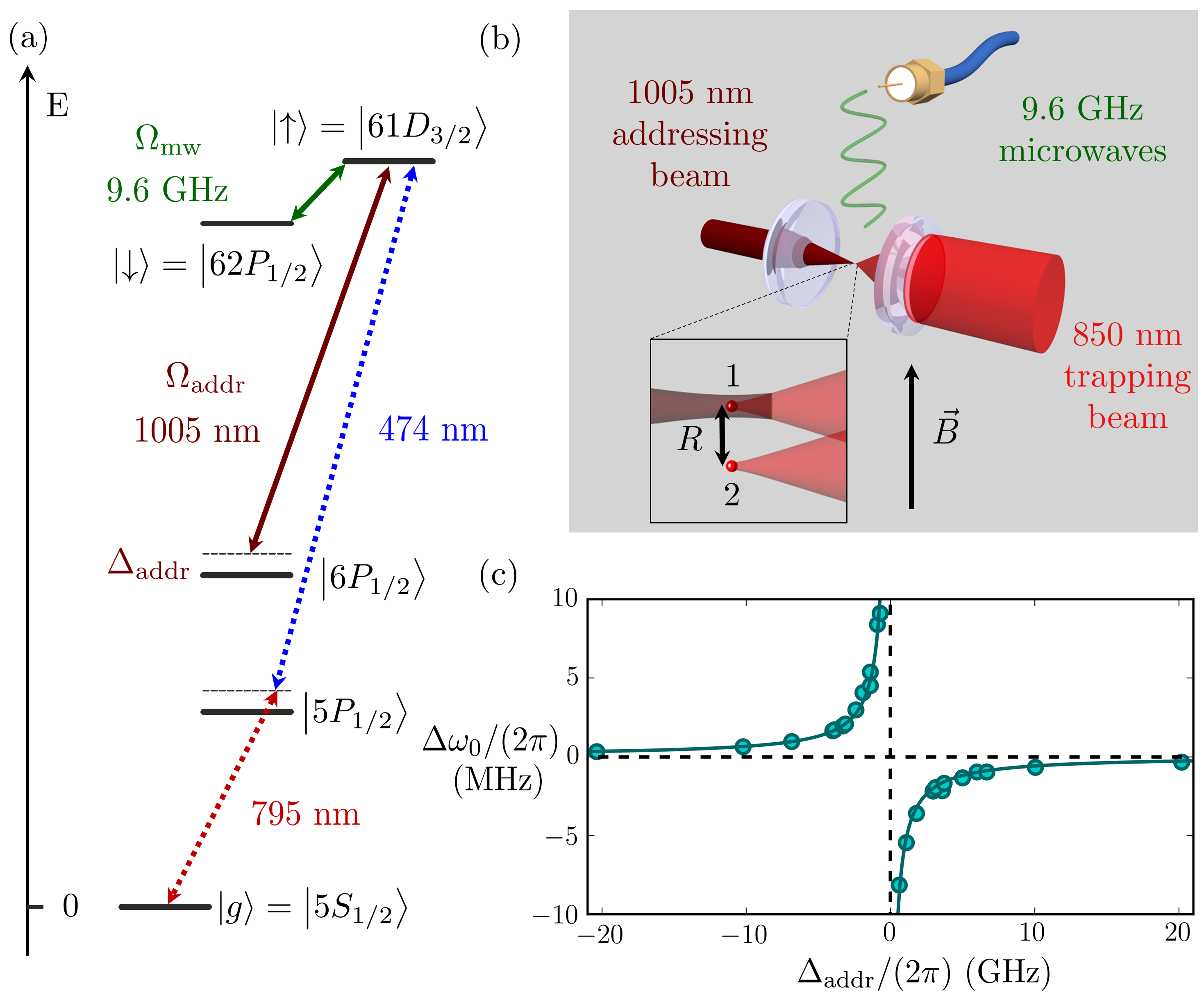}
	\caption{(color online) (a) Scheme of the levels relevant for the experiment. The spin $1/2$ states are encoded in two Rydberg levels. Microwaves couple $\ket{\uparrow}$ to $\ket{\downarrow}$ with Rabi frequency $\Omega_{\text{mw}}$. The addressing beam couples  $\ket{\uparrow}$ with $\ket{6P_{1/2}}$ off-resonantly  to induce a local light-shift $\Delta \omega_0$. 
	(b) Experimental setup. Two atoms separated by a distance $R$ and aligned along the quantization axis, defined by a $B = 7$~G magnetic field. The 1005~nm addressing beam is focused on a single atom. (c) Light-shift $\Delta \omega_0$ of state $\ket{\uparrow}$ measured by microwave spectroscopy, as a function of the detuning $\Delta_{\rm addr}$ of the addressing laser. The full line represents the parameter-free expected light-shift $\Delta \omega_0 = \Omega_{\text{addr}}^2 / 4 \Delta_{\text{addr}}$ with the calculated Rabi frequency $\Omega_{\text{addr}}/(2\pi) = 158$~MHz. Error bars are smaller than the symbols size.} 
	\label{fig:fig1}
\end{figure}

In this Letter, using the Rydberg states $\ket{\uparrow} = \ket{nD}$ and  $\ket{\downarrow} = \ket{n'P}$ of $^{87}$Rb separated by an energy $\hbar\omega_0$ to encode a spin $1/2$, we demonstrate  selective addressing by using a focused addressing laser beam at 1005~nm [see Fig. \ref{fig:fig1}(a)] to induce a controllable light-shift $\Delta\omega_0$ on $\ket{\uparrow}$, while the state $\ket{\downarrow}$ is unaffected. We illustrate this local tuning in a minimal system of two atoms separated by a distance $R$ and governed by the Hamiltonian
\begin{equation}
H = H_0 + \hbar \Delta \omega_0 \frac{\sigma_1^z+1}{2}+ U (\sigma_1^+ \sigma_2^- + \sigma_1^- \sigma_2^+) .
\end{equation}
Here, $H_0 = \hbar \omega_0 (\sigma^z_1 + \sigma^z_2)/2$ is the single-atom Hamiltonian, $U=C_3/R^3$ is the strength of the dipole-dipole interaction, and $\sigma^+_i$, $\sigma^-_i$ and $\sigma^z_i$ denote the spin matrices acting on atom $i =  1, 2$. We first perform microwave spectroscopy on this interacting two-atom system for various addressing light-shifts $\Delta \omega_0$ on atom 1. We show that the two atoms can be brought in and out of the resonant dipole-dipole regime in a controlled way. Notably, in the resonant regime, we observe the collectively enhanced microwave coupling by a factor $\sqrt{2}$ between the state $\ket{\uparrow \uparrow}$ and the ``bright'' or superradiant, superposition of the two atoms $\frac{1}{\sqrt{2}}(\ket{\uparrow \downarrow} + \ket{\downarrow \uparrow})$ as well as the Rydberg blockade inhibiting the spin-flip of the two atoms~\cite{Maxwell2014}. We then demonstrate that the spin-exchange dynamics observed after preparing the initial state $\ket{\uparrow \downarrow}$ can be stopped for a controlled amount of time. A dynamical phase is then accumulated by the addressed atom, due to the energy shift $\hbar \Delta \omega_0$, giving access to various two-atoms entangled states. We show that coherence is maintained by letting the spin-exchange resume. Finally, we discuss the possible limitations of this new tool for future quantum simulation experiments.

The experimental setup, shown schematically in Fig.~\ref{fig:fig1}(b), is described in details in Ref.~\cite{Beguin2013,Labuhn2016}. Briefly, we focus a red-detuned dipole trap beam with an aspheric lens (${\rm NA} = 0.5$) into a MOT of $^{87}$Rb atoms, to a waist of approximately $1.1\,\mu$m.  Multiple traps at arbitrary distances are created by imprinting an appropriate phase on the dipole trap beam (850~nm) with a spatial light modulator prior to focusing~\cite{Nogrette2014}. Single atoms are loaded in the desired traps by active sorting~\cite{Barredo2016}. The temperature of the trapped single atoms is approximately 30~$\mu$K. An external magnetic field of 7~G in the vertical direction defines the quantization axis. 

We choose the Rydberg levels $\ket{\uparrow} = \ket{61D_{3/2},m_j=3/2}$ and $\ket{\downarrow} = \ket{62P_{1/2},m_j=1/2}$ to define the spin $1/2$ [see Fig. \ref{fig:fig1}(a)]. The dipole-dipole coupling between $\ket{\uparrow \downarrow}$ and $\ket{\downarrow \uparrow}$ is $U = C_3 / R^3$ with a calculated $C_3 = h \times 7456\,\text{MHz}.\mu\text{m}^3$~\cite{Sibalic2016,Weber2016} for this choice of Rydberg states. The atoms are excited from the ground state $\ket{g}=\ket{5S_{1/2},F=2,m_F=2}$ to $\ket{\uparrow}$ with a two-photon transition of effective Rabi frequency $4$~MHz. The spin-flip transition $\ket{\uparrow} \leftrightarrow \ket{\downarrow}$ is driven by a resonant microwave pulse at frequency $\omega_0/(2\pi) \simeq 9.600$~GHz emitted by a dipole antenna placed outside the vacuum chamber. At the end of the experiment, a Rydberg de-excitation pulse transfers back atoms in $\ket{\uparrow}$ to the ground state and leaves atoms in $\ket{\downarrow}$ unaffected, which allows selective detection of $\ket{\uparrow}$ and $\ket{\downarrow}$. 

For selective addressing, we use a laser beam at 1005~nm, slightly detuned by a quantity $\Delta_{\text{addr}}$ from the the transition $\ket{6P_{1/2}} \leftrightarrow \ket{nD_{3/2}}$ [see Fig.~\ref{fig:fig1}(a)], which induces a light-shift on $\ket{\uparrow}$, while $\ket{\downarrow}$ is not affected due to the electric dipole selection rules. Initially suggested in ~\cite{Saffman2005}, this scheme was used for magic trapping of ground and Rydberg atoms~\cite{Kuzmich2013}. The addressing beam from a cw Ti:sapphire laser is focused on trap 1 with a linear polarization perpendicular to the quantization axis. We choose a $3.4\,\mu$m waist~\footnote{The beam waist was determined by using the addressing beam as a single-atom trap. The trap depth is measured by Rydberg spectroscopy \cite{Labuhn2014} and the trapping frequency by parametric heating \cite{Piotrowicz2013}.} as a trade-off between adjacent sites cross-talk (1\% residual light-shift at $R = 5.2 \, \mu$m) and alignment issues. The addressing laser is switched on and off by an electro-optic modulator with a rise time of 10~ns. The laser frequency is locked on a commercial wavelength-meter~\footnote{High Finesse, WLM SU10} to prevent long term drifts of $\Delta_{\text{addr}}$. The $\sigma^+$ polarization component of the addressing beam couples $\ket{6P_{1/2}, F = 2, m_F = 2}$ with the Rydberg state $\ket{\uparrow}$, see Fig.~\ref{fig:fig1}(a). The Rabi frequency is calculated~\cite{Sibalic2016} to be $ \Omega_{\text{addr}}/(2\pi) = 158$~MHz for an incident power $P = 30$~mW (only half of the linearly polarized laser power contributes). For a large detuning $\Delta_{\text{addr}} \gg \Omega_{\text{addr}}$, the state $\ket{\uparrow}$ experiences an AC Stark shift $\Delta \omega_0 = \Omega_{\text{addr}}^2 / 4 \Delta_{\text{addr}} $, while the  other state $\ket{\downarrow}$ remains unaffected by this laser (except for a calculated $\sim 4$~MHz ponderomotive light-shift common on both states~\cite{Saffman2005, Raithel2010}). 

In Fig.~\ref{fig:fig1}(c) we present the shifted qubit transition energy $\hbar(\omega_0 + \Delta \omega_0)$ measured by microwave spectroscopy (driving the spin-flip transition) for different detuning $\Delta_{\text{addr}}$ of the addressing beam. The data are in excellent agreement with the expected light-shift using the calculated Rabi frequency and where only the $\ket{6P_{1/2}} \leftrightarrow \ket{61D_{3/2}}$ transition frequency is a free parameter and is measured at $298.139450$~THz. 
This differential light-shift between the two Rydberg states $\ket{\uparrow}$ and $\ket{\downarrow}$ allows tuning the spin-flip transition frequency of atom 1 and bringing it in and out of resonance with atom 2.

\begin{figure}[t]
	\centering
	\includegraphics[width=\linewidth]{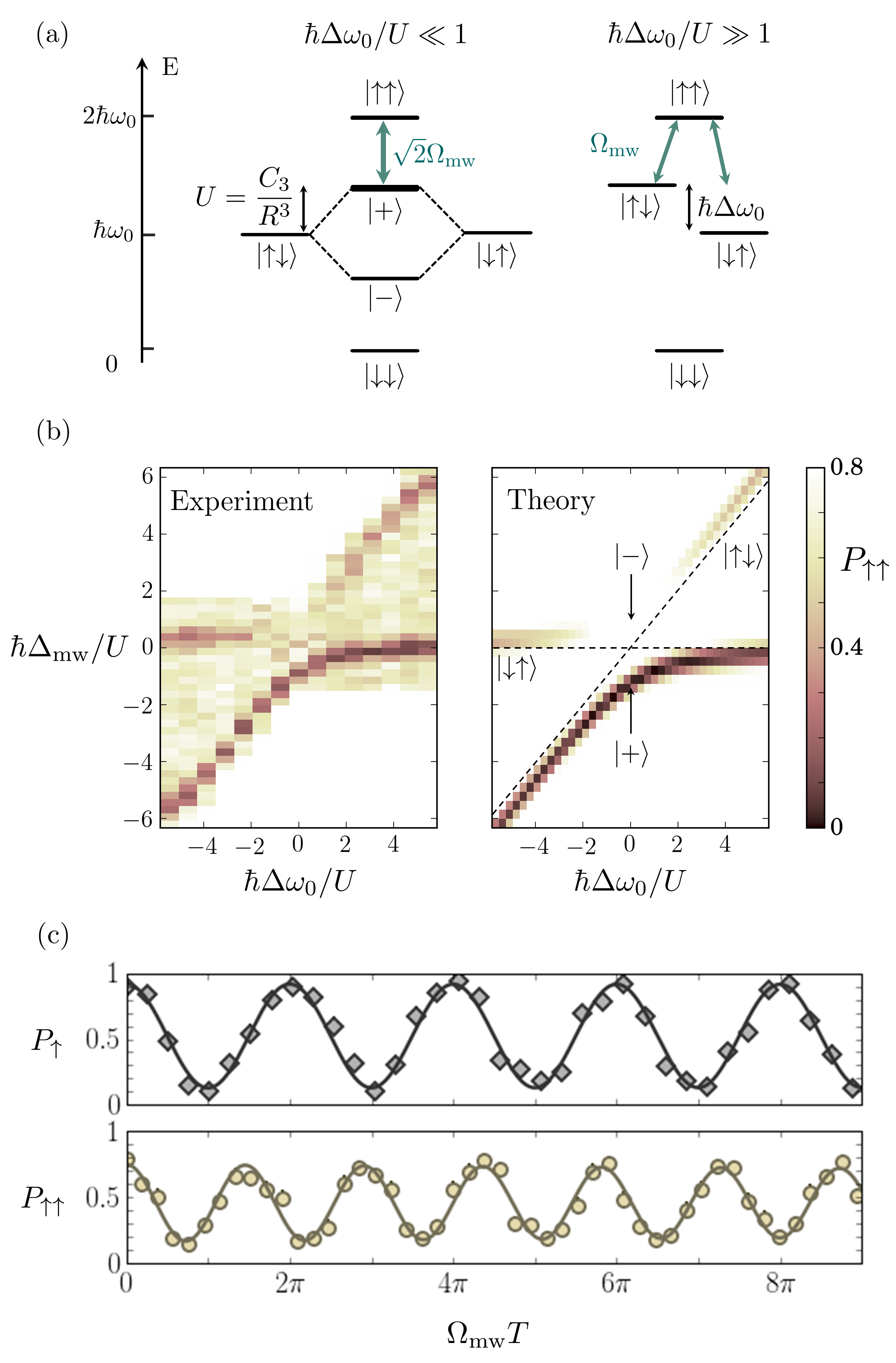}
	\caption{(color online) (a) Two-atom energy structure in the resonant (left) and off-resonant (right) limit. (b) Microwave spectroscopy for $U/h = 0.40 $~MHz ($R=25\,\mu$m) and $\Omega_{\text{mw}}/(2\pi) = 0.1 $~MHz starting from $\ket{\uparrow \uparrow}$. At resonance, the dark state is not coupled anymore by the microwave field. (c) Rabi oscillation for $U/h = 4.09$~MHz ($R = 12 \, \mu$m) driven by microwaves with a Rabi frequency $\Omega_{\text{mw}}/(2\pi) = 1.6$~MHz. Upper panel : driving the single atom transition $\ket{\uparrow} \leftrightarrow \ket{\downarrow}$ at $\Delta_{\text{mw}} = 0$. Lower panel : observing the enhanced microwave coupling on the transition $\ket{\uparrow \uparrow} \leftrightarrow \ket{+}$ at $\Delta_{\text{mw}} = -U$. Error bars are smaller than the symbols size.} 
	\label{fig:fig2}
\end{figure}

The effect of the dipole-dipole interaction between two atoms can now be controlled with the addressing beam. We first recall the two-atom energy spectrum in the presence of the dipole-dipole interaction $U$ and the addressing light-shift $ \hbar \Delta \omega_0$ as illustrated in Fig.~\ref{fig:fig2}(a). Without the addressing laser, the dipole-dipole interaction lifts the degeneracy between $\ket{\uparrow \downarrow}$ and $\ket{\downarrow \uparrow}$ and the Hamiltonian eigenstates become $\ket{\pm} = \frac{1}{\sqrt{2}}(\ket{\uparrow \downarrow} \pm \ket{\downarrow \uparrow})$ separated in energy by $2 U = 2 C_3 /R^3$. With the addressing laser focused on the first atom, the state $\ket{\uparrow \downarrow}$ is shifted by an energy $\hbar \Delta \omega_0$ and is not resonant anymore with $\ket{\downarrow \uparrow}$. A microwave field at angular frequency $\omega = \omega_0 + \Delta_{\text{mw}}$ drives single spin-flip transitions with Rabi frequency $\Omega_{\text{mw}}$ to probe the system. At resonance ($\Delta \omega_0 = 0$), the microwave coupling between the superradiant state $\ket{+}$ and $\ket{\uparrow \uparrow}$ is enhanced to $\sqrt{2} \Omega_{\text{mw}}$ while it is suppressed for the subradiant state $\ket{-}$. In the limit of strong addressing $\hbar \Delta \omega_0 \gg U$, the dipole-dipole interaction can be neglected and the microwave field is expected to couple equally $\ket{\uparrow \uparrow}$ to the two single-flipped spin states. 

To study the system by microwave spectroscopy, the experiment starts with the two atoms separated by $R = 25\,\mu$m, aligned along the quantization axis, and prepared in the state $\ket{\uparrow \uparrow}$ after the initial Rydberg excitation pulse. The addressing laser and microwave field ($\Omega_{\text{mw}}/(2\pi) = 0.1 $~MHz) are then switched on during a time $\tau = \pi/(\sqrt{2} \Omega_{\text{mw}}) = 3.5 \, \mu$s to induce a spin-flip before a final Rydberg de-excitation pulse allows the read-out of the final population in $\ket{\uparrow \uparrow}$. The experiment is repeated $\approx 100$ times for each set of parameters $\Delta_{\text{mw}}, \Delta \omega_0$ (the latter being tuned by changing the addressing beam detuning $\Delta_{\text{addr}}$) and the results are shown in Fig.~\ref{fig:fig2}(b). The microwave resonances, seen as a drop in the final population of state $\ket{\uparrow \uparrow}$, are very well reproduced by calculations without any adjustable parameters (right panel). Notably at $\Delta \omega_0 = 0$, the ``dark'' (subradiant) state $\ket{-}$ is not coupled anymore and the bright state $\ket{+}$ energy shift measures the interaction strength $U/h = 0.40$~MHz.

The enhanced microwave coupling to the ``bright'' (superradiant) state is best seen on the Rabi oscillation in Fig.~\ref{fig:fig2}(c). The upper panel shows a single-atom Rabi oscillation between states $\ket{\uparrow}$ and $\ket{\downarrow}$ driven at $\Omega_{\text{mw}}/(2\pi) = 1.6$~MHz. The lower panel shows coherent oscillations between  $\ket{\uparrow \uparrow}$ and $\ket{+}$, with a measured frequency enhancement of $1.375(5)$, close to the expected $\sqrt{2}$. The finite contrast of these microwave-driven Rabi oscillations is due to the finite efficiency $\eta = 0.88$ of the Rydberg excitation pulse, resulting directly in imperfect initial transfer of each atom in $\ket{\uparrow}$ and in small measurement errors~\cite{Labuhn2016}. In this experiment, the state $\ket{\downarrow \downarrow}$ is not populated as a consequence of the Rydberg blockade shifting the singly excited state $\ket{+}$.

\begin{figure}
\centering
\includegraphics[width=\linewidth]{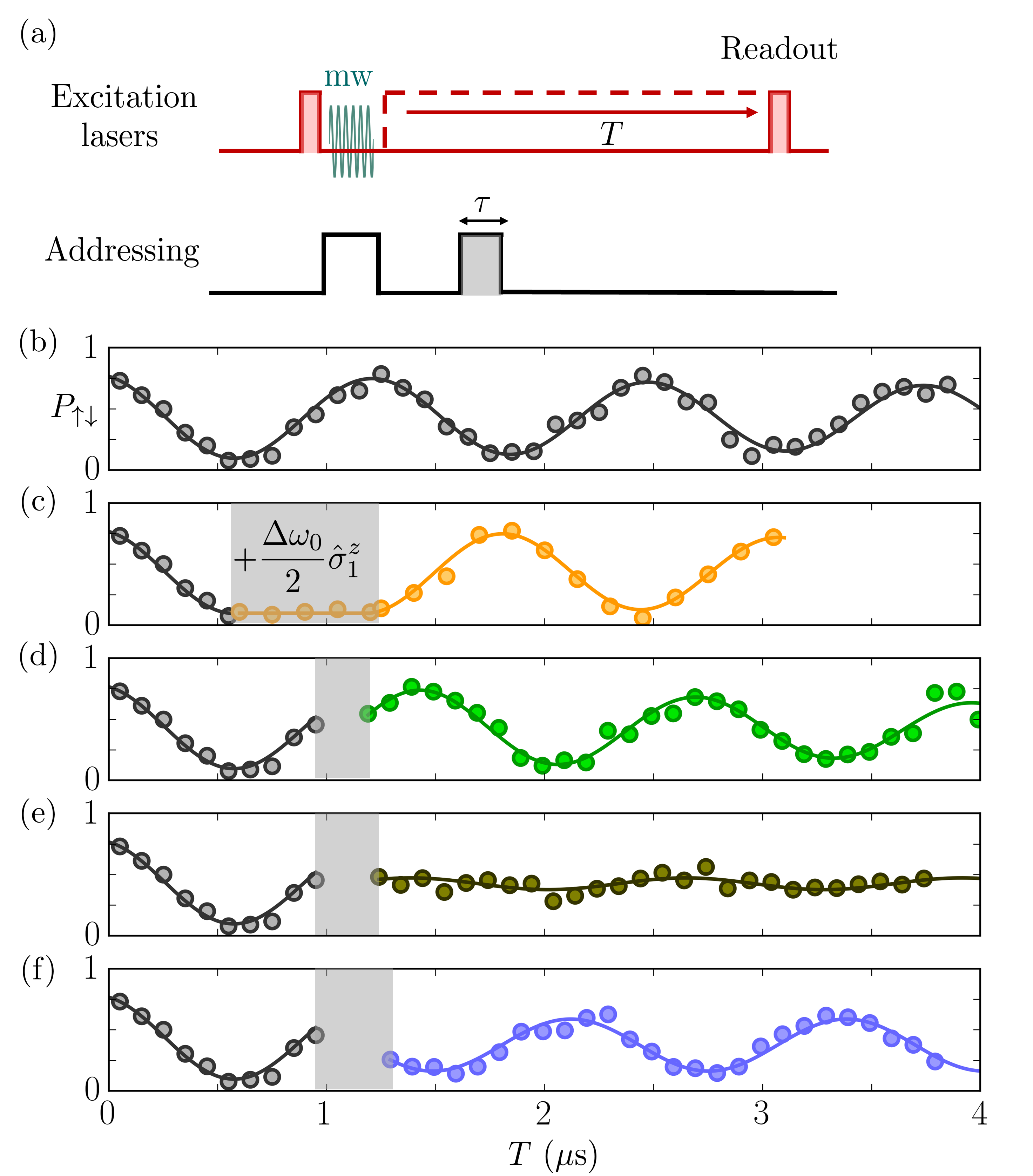}
\caption{(color online) (a) Experimental sequence to observe the freezing of the spin-exchange dynamics by the addressing beam with $\Delta \omega_0 /(2\pi) = 4.8$~MHz and $U/h = 0.40$~MHz ($R=25\,\mu$m). (b) Spin-exchange dynamics $\ket{\uparrow \downarrow} \leftrightarrow \ket{\downarrow \uparrow}$ driven by the dipole-dipole interaction at a measured frequency $2U/h = 0.80$~MHz. (c) The addressing beam stops the exchange during a time $\tau = 600$~ns (gray area). (d-f) : the spin-exchange is frozen when $\ket{\psi} = -\frac{1}{\sqrt{2}}(\ket{\uparrow \downarrow} + i \ket{\downarrow \uparrow})$ and the addressing energy shift imprints a relative dynamical phase $\Delta \omega_0 \tau = 2\pi, 2.5 \pi, 3 \pi$ (d,e,f) between $\ket{\uparrow \downarrow}$ and $\ket{\downarrow \uparrow}$. Error bars are smaller than the symbols size. Solid lines are sinusoidal fits with fixed frequency $2U/h$.} 
\label{fig:fig3}
\end{figure}

We now use the addressing beam as a tool to freeze at will the spin-exchange dynamics between the two states $\ket{\uparrow \downarrow}$ and $\ket{\downarrow \uparrow}$. The experimental sequence is shown in Fig.~\ref{fig:fig3}(a) and starts as previously with two atoms excited in $\ket{\uparrow \uparrow}$. The atoms are then initialized in $\ket{\uparrow \downarrow}$ by addressing atom 1  with $\Delta \omega_0 /(2\pi) = 4.8$~MHz, while a global microwave pulse ($\Omega_{\text{mw}}/(2\pi) = 1.3$~MHz) only drives the spin-flip transition on atom 2. Once the addressing laser is switched off, the system evolves for a time $T$ before the read-out pulse and shows coherent, interaction-driven spin-exchange dynamics at a measured frequency $2U/h = 0.80$~MHz as observed on the state population $P_{\uparrow \downarrow}$ in Fig.~\ref{fig:fig3}(b). After half a period of spin exchange, we shine again the addressing beam during a time $\tau = 600$~ns (gray area in Fig.~\ref{fig:fig3}(c)) thus detuning the first atom out of resonance ($\hbar \Delta \omega_0 \gg U$). The system is then frozen in state $\ket{\downarrow \uparrow}$, until the dynamic restarts without any noticeable loss of contrast after the addressing beam is turned off.

During the freezing time, the energy-shifted state $\ket{\uparrow \downarrow}$ acquires a dynamical phase $\phi = \Delta \omega_0 \tau$ compared to $\ket{\downarrow \uparrow}$. To observe this relative phase, we switch on the addressing beam when the system is in the superposition of states $-\frac{1}{\sqrt{2}}(\ket{\uparrow \downarrow} + i \ket{\downarrow \uparrow})$. During the addressing time $\tau$ it evolves into $\ket{\psi} = -\frac{1}{\sqrt{2}}(e^{-i\phi}\ket{\uparrow \downarrow} + i \ket{\downarrow \uparrow})$ and the following spin-exchange dynamics depends on the acquired phase $\phi$. On Fig.~\ref{fig:fig3}(d), we adjust the addressing time $\tau$ such that $\phi = 2 \pi$ and the dynamics resumes as before we froze it. For $\phi = 2.5 \pi$, $\ket{\psi} = i\ket{-}$ is an eigenstate of the Hamiltonian and thus the populations do not evolve anymore as seen in Fig.~\ref{fig:fig3}(e). Finally in Fig.~\ref{fig:fig3}(f), we obtain a $\pi$-phase shift on the spin-exchange dynamics when $\phi = 3 \pi$. This illustrates the potential of the addressing beam for coherent manipulation of many-body quantum states.

We finally discuss the possible imperfections of this addressing technique for quantum state engineering. A first decoherence mechanism is the spontaneous emission due to the off-resonant coupling to $\ket{6P_{1/2}}$, whose lifetime is $\tau_{6P} = 121$~ns~\cite{Gomez2004}, limiting the lifetime of the addressed state to $\tau_{\uparrow} \simeq (\Delta_{\text{addr}}/\Omega_{\text{addr}})^2\, \tau_{6P} $~\cite{Saffman2005}. For our typical parameters, $\tau_\uparrow$ reaches several tens of microseconds, close to the natural lifetime of the involved Rydberg levels, and thus barely affects the dynamics of the system. Another process can also bring the atom out of the spin $1/2$ basis : due to the $\sigma^+ + \sigma^-$ polarization of the addressing beam, both Zeeman states $\ket{\uparrow} = \ket{nD_{3/2}, m_J = 3/2}$ and $\ket{0} = \ket{nD_{3/2}, m_J = -1/2}$ are coupled to $\ket{6P_{1/2}}$. An atom in state $\ket{\uparrow}$ can thus be transferred to $\ket{0}$ in a Raman process with effective Rabi frequency $\Omega_{\text{Raman}} = \frac{1}{\sqrt{3}}\Omega_\text{addr}^2/(2\Delta_{\text{addr}}) = \frac{2}{\sqrt{3}} \Delta \omega_0$. Due to the magnetic field, the two Zeeman levels $\ket{\uparrow}$ and $\ket{0}$ are separated in frequency by $\delta/(2\pi) = 15$~MHz and the Raman process is suppressed as long as $\Omega_{\text{Raman}} \ll \delta$. Simulations show that for the data presented here, this process gives a transfer probability of at most $\sim 8$\%. Using a higher magnetic field, splitting even more the Zeeman structure, would be a straightforward way to reduce this transfer out of the qubit subspace. Finally, for some applications in quantum state engineering, it is important to have a fine control of the dynamical phase $\phi =\Delta \omega_0 \tau$, which requires to limit  shot-to-shot fluctuations of the addressing beam power. All the above imperfections can be made negligible by modest technical improvements.

In summary, we have demonstrated local selective shifts of Rydberg levels, allowing us to bring two atoms in and out of the resonant dipole-dipole interaction regime. This makes it possible to create entangled states of two atoms in different ways. A first possibility consists in starting from $\ket{\uparrow \uparrow}$ and in collectively driving the system with microwaves to $\ket{+}$ (Fig.~\ref{fig:fig2}). In contrast to the usual Rydberg blockade protocols, based on optical driving between $\ket{g}$ and $\ket{\uparrow}$, this approach benefits from (i) the high amplitude and phase stability of microwave sources, (ii) the long wavelength of microwave fields compared to the interatomic spacing, making motional phases negligible~\cite{Wilk2010} and (iii) the fact that the entangled state $\ket{+}$ is separated by a finite energy gap from $\ket{-}$, thus decreasing its  sensitivity to dephasing~\cite{Viscor2017}. A second possibility is to start from $\ket{\uparrow \downarrow}$ and to stop the spin-exchange dynamics at the appropriate time (Fig.~\ref{fig:fig3}), which allows to create any coherent superposition of $\ket{\uparrow \downarrow}$ and $\ket{\downarrow \uparrow}$, in particular the otherwise inaccessible ``dark'' subradiant state $\ket{-}$. In future work, one could even generate several independently controlled addressing beams using spatial light modulators. These results open exciting prospects for quantum state engineering, quantum tomography, and quantum simulation of the XY model using arrays of single Rydberg atoms~\cite{Zoller2017}.

This  work  benefited  from  financial  support  by  the  EU  [H2020 FET-PROACT  Project RySQ],  by  the  PALM  Labex  (projects  QUANTICA and XYLOS)  and  by the R\'egion \^Ile-de-France in the framework of DIM Nano-K.

$^{\ast}$ S. de L., D. B. and V. L. contributed equally to this work. 

\bibliography{de_Leseleuc_refs}

\begin{thebibliography}{46}%
\makeatletter
\providecommand \@ifxundefined [1]{%
 \@ifx{#1\undefined}
}%
\providecommand \@ifnum [1]{%
 \ifnum #1\expandafter \@firstoftwo
 \else \expandafter \@secondoftwo
 \fi
}%
\providecommand \@ifx [1]{%
 \ifx #1\expandafter \@firstoftwo
 \else \expandafter \@secondoftwo
 \fi
}%
\providecommand \natexlab [1]{#1}%
\providecommand \enquote  [1]{``#1''}%
\providecommand \bibnamefont  [1]{#1}%
\providecommand \bibfnamefont [1]{#1}%
\providecommand \citenamefont [1]{#1}%
\providecommand \href@noop [0]{\@secondoftwo}%
\providecommand \href [0]{\begingroup \@sanitize@url \@href}%
\providecommand \@href[1]{\@@startlink{#1}\@@href}%
\providecommand \@@href[1]{\endgroup#1\@@endlink}%
\providecommand \@sanitize@url [0]{\catcode `\\12\catcode `\$12\catcode
  `\&12\catcode `\#12\catcode `\^12\catcode `\_12\catcode `\%12\relax}%
\providecommand \@@startlink[1]{}%
\providecommand \@@endlink[0]{}%
\providecommand \url  [0]{\begingroup\@sanitize@url \@url }%
\providecommand \@url [1]{\endgroup\@href {#1}{\urlprefix }}%
\providecommand \urlprefix  [0]{URL }%
\providecommand \Eprint [0]{\href }%
\providecommand \doibase [0]{http://dx.doi.org/}%
\providecommand \selectlanguage [0]{\@gobble}%
\providecommand \bibinfo  [0]{\@secondoftwo}%
\providecommand \bibfield  [0]{\@secondoftwo}%
\providecommand \translation [1]{[#1]}%
\providecommand \BibitemOpen [0]{}%
\providecommand \bibitemStop [0]{}%
\providecommand \bibitemNoStop [0]{.\EOS\space}%
\providecommand \EOS [0]{\spacefactor3000\relax}%
\providecommand \BibitemShut  [1]{\csname bibitem#1\endcsname}%
\let\auto@bib@innerbib\@empty
%</preamble>
\bibitem [{\citenamefont {Georgescu}\ \emph {et~al.}(2014)\citenamefont
  {Georgescu}, \citenamefont {Ashhab},\ and\ \citenamefont
  {Nori}}]{Georgescu2014}%
  \BibitemOpen
  \bibfield  {author} {\bibinfo {author} {\bibfnamefont {I.~M.}\ \bibnamefont
  {Georgescu}}, \bibinfo {author} {\bibfnamefont {S.}~\bibnamefont {Ashhab}}, \
  and\ \bibinfo {author} {\bibfnamefont {F.}~\bibnamefont {Nori}},\ }\href
  {\doibase 10.1103/RevModPhys.86.153} {\bibfield  {journal} {\bibinfo
  {journal} {Rev. Mod. Phys.}\ }\textbf {\bibinfo {volume} {86}},\ \bibinfo
  {pages} {153} (\bibinfo {year} {2014})}\BibitemShut {NoStop}%
\bibitem [{\citenamefont {Blatt}\ and\ \citenamefont
  {Wineland}(2008)}]{Blatt2008}%
  \BibitemOpen
  \bibfield  {author} {\bibinfo {author} {\bibfnamefont {R.}~\bibnamefont
  {Blatt}}\ and\ \bibinfo {author} {\bibfnamefont {D.}~\bibnamefont
  {Wineland}},\ }\href {\doibase 10.1038/nature07125} {\bibfield  {journal}
  {\bibinfo  {journal} {Nature}\ }\textbf {\bibinfo {volume} {453}},\ \bibinfo
  {pages} {1008} (\bibinfo {year} {2008})}\BibitemShut {NoStop}%
\bibitem [{\citenamefont {Veldhorst}\ \emph {et~al.}(2015)\citenamefont
  {Veldhorst}, \citenamefont {Yang}, \citenamefont {Hwang}, \citenamefont
  {Huang}, \citenamefont {Dehollain}, \citenamefont {Muhonen}, \citenamefont
  {Simmons}, \citenamefont {Laucht}, \citenamefont {Hudson}, \citenamefont
  {Itoh}, \citenamefont {Morello},\ and\ \citenamefont
  {Dzurak}}]{Veldhorst2015}%
  \BibitemOpen
  \bibfield  {author} {\bibinfo {author} {\bibfnamefont {M.}~\bibnamefont
  {Veldhorst}}, \bibinfo {author} {\bibfnamefont {C.~H.}\ \bibnamefont {Yang}},
  \bibinfo {author} {\bibfnamefont {J.~C.~C.}\ \bibnamefont {Hwang}}, \bibinfo
  {author} {\bibfnamefont {W.}~\bibnamefont {Huang}}, \bibinfo {author}
  {\bibfnamefont {J.~P.}\ \bibnamefont {Dehollain}}, \bibinfo {author}
  {\bibfnamefont {J.~T.}\ \bibnamefont {Muhonen}}, \bibinfo {author}
  {\bibfnamefont {S.}~\bibnamefont {Simmons}}, \bibinfo {author} {\bibfnamefont
  {A.}~\bibnamefont {Laucht}}, \bibinfo {author} {\bibfnamefont {F.~E.}\
  \bibnamefont {Hudson}}, \bibinfo {author} {\bibfnamefont {K.~M.}\
  \bibnamefont {Itoh}}, \bibinfo {author} {\bibfnamefont {A.}~\bibnamefont
  {Morello}}, \ and\ \bibinfo {author} {\bibfnamefont {A.~S.}\ \bibnamefont
  {Dzurak}},\ }\href {\doibase 10.1038/nature15263} {\bibfield  {journal}
  {\bibinfo  {journal} {Nature}\ }\textbf {\bibinfo {volume} {526}},\ \bibinfo
  {pages} {410} (\bibinfo {year} {2015})}\BibitemShut {NoStop}%
\bibitem [{\citenamefont {Salath{\'{e}}}\ \emph {et~al.}(2015)\citenamefont
  {Salath{\'{e}}}, \citenamefont {Mondal}, \citenamefont {Oppliger},
  \citenamefont {Heinsoo}, \citenamefont {Kurpiers}, \citenamefont
  {Poto{\v{c}}nik}, \citenamefont {Mezzacapo}, \citenamefont {{Las Heras}},
  \citenamefont {Lamata}, \citenamefont {Solano}, \citenamefont {Filipp},\ and\
  \citenamefont {Wallraff}}]{Salathe2015}%
  \BibitemOpen
  \bibfield  {author} {\bibinfo {author} {\bibfnamefont {Y.}~\bibnamefont
  {Salath{\'{e}}}}, \bibinfo {author} {\bibfnamefont {M.}~\bibnamefont
  {Mondal}}, \bibinfo {author} {\bibfnamefont {M.}~\bibnamefont {Oppliger}},
  \bibinfo {author} {\bibfnamefont {J.}~\bibnamefont {Heinsoo}}, \bibinfo
  {author} {\bibfnamefont {P.}~\bibnamefont {Kurpiers}}, \bibinfo {author}
  {\bibfnamefont {A.}~\bibnamefont {Poto{\v{c}}nik}}, \bibinfo {author}
  {\bibfnamefont {A.}~\bibnamefont {Mezzacapo}}, \bibinfo {author}
  {\bibfnamefont {U.}~\bibnamefont {{Las Heras}}}, \bibinfo {author}
  {\bibfnamefont {L.}~\bibnamefont {Lamata}}, \bibinfo {author} {\bibfnamefont
  {E.}~\bibnamefont {Solano}}, \bibinfo {author} {\bibfnamefont
  {S.}~\bibnamefont {Filipp}}, \ and\ \bibinfo {author} {\bibfnamefont
  {A.}~\bibnamefont {Wallraff}},\ }\href {\doibase 10.1103/PhysRevX.5.021027}
  {\bibfield  {journal} {\bibinfo  {journal} {Phys. Rev. X}\ }\textbf {\bibinfo
  {volume} {5}},\ \bibinfo {pages} {1} (\bibinfo {year} {2015})}\BibitemShut
  {NoStop}%
\bibitem [{\citenamefont {Choi}\ \emph {et~al.}(2016)\citenamefont {Choi},
  \citenamefont {Hild}, \citenamefont {Zeiher}, \citenamefont {Schauss},
  \citenamefont {Rubio-Abadal}, \citenamefont {Yefsah}, \citenamefont
  {Khemani}, \citenamefont {Huse}, \citenamefont {Bloch},\ and\ \citenamefont
  {Gross}}]{Choi2016}%
  \BibitemOpen
  \bibfield  {author} {\bibinfo {author} {\bibfnamefont {J.-Y.}\ \bibnamefont
  {Choi}}, \bibinfo {author} {\bibfnamefont {S.}~\bibnamefont {Hild}}, \bibinfo
  {author} {\bibfnamefont {J.}~\bibnamefont {Zeiher}}, \bibinfo {author}
  {\bibfnamefont {P.}~\bibnamefont {Schauss}}, \bibinfo {author} {\bibfnamefont
  {A.}~\bibnamefont {Rubio-Abadal}}, \bibinfo {author} {\bibfnamefont
  {T.}~\bibnamefont {Yefsah}}, \bibinfo {author} {\bibfnamefont
  {V.}~\bibnamefont {Khemani}}, \bibinfo {author} {\bibfnamefont {D.~A.}\
  \bibnamefont {Huse}}, \bibinfo {author} {\bibfnamefont {I.}~\bibnamefont
  {Bloch}}, \ and\ \bibinfo {author} {\bibfnamefont {C.}~\bibnamefont
  {Gross}},\ }\href {\doibase 10.1126/science.aaf8834} {\bibfield  {journal}
  {\bibinfo  {journal} {Science}\ }\textbf {\bibinfo {volume} {352}},\ \bibinfo
  {pages} {1547} (\bibinfo {year} {2016})}\BibitemShut {NoStop}%
\bibitem [{\citenamefont {Marcuzzi}\ \emph {et~al.}(2017)\citenamefont
  {Marcuzzi}, \citenamefont {Min{\'{a}}{\v{r}}}, \citenamefont {Barredo},
  \citenamefont {de~L{\'{e}}s{\'{e}}leuc}, \citenamefont {Labuhn},
  \citenamefont {Lahaye}, \citenamefont {Browaeys}, \citenamefont {Levi},\ and\
  \citenamefont {Lesanovsky}}]{Marcuzzi2017}%
  \BibitemOpen
  \bibfield  {author} {\bibinfo {author} {\bibfnamefont {M.}~\bibnamefont
  {Marcuzzi}}, \bibinfo {author} {\bibfnamefont {J.}~\bibnamefont
  {Min{\'{a}}{\v{r}}}}, \bibinfo {author} {\bibfnamefont {D.}~\bibnamefont
  {Barredo}}, \bibinfo {author} {\bibfnamefont {S.}~\bibnamefont
  {de~L{\'{e}}s{\'{e}}leuc}}, \bibinfo {author} {\bibfnamefont
  {H.}~\bibnamefont {Labuhn}}, \bibinfo {author} {\bibfnamefont
  {T.}~\bibnamefont {Lahaye}}, \bibinfo {author} {\bibfnamefont
  {A.}~\bibnamefont {Browaeys}}, \bibinfo {author} {\bibfnamefont
  {E.}~\bibnamefont {Levi}}, \ and\ \bibinfo {author} {\bibfnamefont
  {I.}~\bibnamefont {Lesanovsky}},\ }\href {\doibase
  10.1103/PhysRevLett.118.063606} {\bibfield  {journal} {\bibinfo  {journal}
  {Phys. Rev. Lett.}\ }\textbf {\bibinfo {volume} {118}},\ \bibinfo {pages}
  {063606} (\bibinfo {year} {2017})}\BibitemShut {NoStop}%
\bibitem [{\citenamefont {Smith}\ \emph {et~al.}(2015)\citenamefont {Smith},
  \citenamefont {Lee}, \citenamefont {Richerme}, \citenamefont {Neyenhuis},
  \citenamefont {Hess}, \citenamefont {Hauke}, \citenamefont {Heyl},
  \citenamefont {Huse},\ and\ \citenamefont {Monroe}}]{Smith2015}%
  \BibitemOpen
  \bibfield  {author} {\bibinfo {author} {\bibfnamefont {J.}~\bibnamefont
  {Smith}}, \bibinfo {author} {\bibfnamefont {A.}~\bibnamefont {Lee}}, \bibinfo
  {author} {\bibfnamefont {P.}~\bibnamefont {Richerme}}, \bibinfo {author}
  {\bibfnamefont {B.}~\bibnamefont {Neyenhuis}}, \bibinfo {author}
  {\bibfnamefont {P.~W.}\ \bibnamefont {Hess}}, \bibinfo {author}
  {\bibfnamefont {P.}~\bibnamefont {Hauke}}, \bibinfo {author} {\bibfnamefont
  {M.}~\bibnamefont {Heyl}}, \bibinfo {author} {\bibfnamefont {D.~A.}\
  \bibnamefont {Huse}}, \ and\ \bibinfo {author} {\bibfnamefont
  {C.}~\bibnamefont {Monroe}},\ }\href {\doibase 10.1038/nphys3783} {\bibfield
  {journal} {\bibinfo  {journal} {Nature Phys.}\ }\textbf {\bibinfo {volume}
  {12}},\ \bibinfo {pages} {907} (\bibinfo {year} {2015})}\BibitemShut
  {NoStop}%
\bibitem [{\citenamefont {Veldhorst}\ \emph {et~al.}(2014)\citenamefont
  {Veldhorst}, \citenamefont {Hwang}, \citenamefont {Yang}, \citenamefont
  {Leenstra}, \citenamefont {de~Ronde}, \citenamefont {Dehollain},
  \citenamefont {Muhonen}, \citenamefont {Hudson}, \citenamefont {Itoh},
  \citenamefont {Morello},\ and\ \citenamefont {Dzurak}}]{Veldhorst2014}%
  \BibitemOpen
  \bibfield  {author} {\bibinfo {author} {\bibfnamefont {M.}~\bibnamefont
  {Veldhorst}}, \bibinfo {author} {\bibfnamefont {J.~C.~C.}\ \bibnamefont
  {Hwang}}, \bibinfo {author} {\bibfnamefont {C.~H.}\ \bibnamefont {Yang}},
  \bibinfo {author} {\bibfnamefont {A.~W.}\ \bibnamefont {Leenstra}}, \bibinfo
  {author} {\bibfnamefont {B.}~\bibnamefont {de~Ronde}}, \bibinfo {author}
  {\bibfnamefont {J.~P.}\ \bibnamefont {Dehollain}}, \bibinfo {author}
  {\bibfnamefont {J.~T.}\ \bibnamefont {Muhonen}}, \bibinfo {author}
  {\bibfnamefont {F.~E.}\ \bibnamefont {Hudson}}, \bibinfo {author}
  {\bibfnamefont {K.~M.}\ \bibnamefont {Itoh}}, \bibinfo {author}
  {\bibfnamefont {A.}~\bibnamefont {Morello}}, \ and\ \bibinfo {author}
  {\bibfnamefont {A.~S.}\ \bibnamefont {Dzurak}},\ }\href {\doibase
  10.1038/nnano.2014.216} {\bibfield  {journal} {\bibinfo  {journal} {Nature
  Nano.}\ }\textbf {\bibinfo {volume} {9}},\ \bibinfo {pages} {981} (\bibinfo
  {year} {2014})}\BibitemShut {NoStop}%
\bibitem [{\citenamefont {Houck}\ \emph {et~al.}(2012)\citenamefont {Houck},
  \citenamefont {T{\"{u}}reci},\ and\ \citenamefont {Koch}}]{Houck2012}%
  \BibitemOpen
  \bibfield  {author} {\bibinfo {author} {\bibfnamefont {A.~A.}\ \bibnamefont
  {Houck}}, \bibinfo {author} {\bibfnamefont {H.~E.}\ \bibnamefont
  {T{\"{u}}reci}}, \ and\ \bibinfo {author} {\bibfnamefont {J.}~\bibnamefont
  {Koch}},\ }\href {\doibase 10.1038/nphys2251} {\bibfield  {journal} {\bibinfo
   {journal} {Nature Phys.}\ }\textbf {\bibinfo {volume} {8}},\ \bibinfo
  {pages} {292} (\bibinfo {year} {2012})}\BibitemShut {NoStop}%
\bibitem [{\citenamefont {Weitenberg}\ \emph {et~al.}(2011)\citenamefont
  {Weitenberg}, \citenamefont {Endres}, \citenamefont {Sherson}, \citenamefont
  {Cheneau}, \citenamefont {Schauss}, \citenamefont {Fukuhara}, \citenamefont
  {Bloch},\ and\ \citenamefont {Kuhr}}]{Weitenberg2011}%
  \BibitemOpen
  \bibfield  {author} {\bibinfo {author} {\bibfnamefont {C.}~\bibnamefont
  {Weitenberg}}, \bibinfo {author} {\bibfnamefont {M.}~\bibnamefont {Endres}},
  \bibinfo {author} {\bibfnamefont {J.~F.}\ \bibnamefont {Sherson}}, \bibinfo
  {author} {\bibfnamefont {M.}~\bibnamefont {Cheneau}}, \bibinfo {author}
  {\bibfnamefont {P.}~\bibnamefont {Schauss}}, \bibinfo {author} {\bibfnamefont
  {T.}~\bibnamefont {Fukuhara}}, \bibinfo {author} {\bibfnamefont
  {I.}~\bibnamefont {Bloch}}, \ and\ \bibinfo {author} {\bibfnamefont
  {S.}~\bibnamefont {Kuhr}},\ }\href {\doibase 10.1038/nature09827} {\bibfield
  {journal} {\bibinfo  {journal} {Nature}\ }\textbf {\bibinfo {volume} {471}},\
  \bibinfo {pages} {319} (\bibinfo {year} {2011})}\BibitemShut {NoStop}%
\bibitem [{\citenamefont {Labuhn}\ \emph {et~al.}(2014)\citenamefont {Labuhn},
  \citenamefont {Ravets}, \citenamefont {Barredo}, \citenamefont {B\'eguin},
  \citenamefont {Nogrette}, \citenamefont {Lahaye},\ and\ \citenamefont
  {Browaeys}}]{Labuhn2014}%
  \BibitemOpen
  \bibfield  {author} {\bibinfo {author} {\bibfnamefont {H.}~\bibnamefont
  {Labuhn}}, \bibinfo {author} {\bibfnamefont {S.}~\bibnamefont {Ravets}},
  \bibinfo {author} {\bibfnamefont {D.}~\bibnamefont {Barredo}}, \bibinfo
  {author} {\bibfnamefont {L.}~\bibnamefont {B\'eguin}}, \bibinfo {author}
  {\bibfnamefont {F.}~\bibnamefont {Nogrette}}, \bibinfo {author}
  {\bibfnamefont {T.}~\bibnamefont {Lahaye}}, \ and\ \bibinfo {author}
  {\bibfnamefont {A.}~\bibnamefont {Browaeys}},\ }\href {\doibase
  10.1103/PhysRevA.90.023415} {\bibfield  {journal} {\bibinfo  {journal} {Phys.
  Rev. A}\ }\textbf {\bibinfo {volume} {90}},\ \bibinfo {pages} {023415}
  (\bibinfo {year} {2014})}\BibitemShut {NoStop}%
\bibitem [{\citenamefont {Xia}\ \emph {et~al.}(2015)\citenamefont {Xia},
  \citenamefont {Lichtman}, \citenamefont {Maller}, \citenamefont {Carr},
  \citenamefont {Piotrowicz}, \citenamefont {Isenhower},\ and\ \citenamefont
  {Saffman}}]{Xia2015}%
  \BibitemOpen
  \bibfield  {author} {\bibinfo {author} {\bibfnamefont {T.}~\bibnamefont
  {Xia}}, \bibinfo {author} {\bibfnamefont {M.}~\bibnamefont {Lichtman}},
  \bibinfo {author} {\bibfnamefont {K.}~\bibnamefont {Maller}}, \bibinfo
  {author} {\bibfnamefont {A.~W.}\ \bibnamefont {Carr}}, \bibinfo {author}
  {\bibfnamefont {M.~J.}\ \bibnamefont {Piotrowicz}}, \bibinfo {author}
  {\bibfnamefont {L.}~\bibnamefont {Isenhower}}, \ and\ \bibinfo {author}
  {\bibfnamefont {M.}~\bibnamefont {Saffman}},\ }\href {\doibase
  10.1103/PhysRevLett.114.100503} {\bibfield  {journal} {\bibinfo  {journal}
  {Phys. Rev. Lett.}\ }\textbf {\bibinfo {volume} {114}},\ \bibinfo {pages}
  {100503} (\bibinfo {year} {2015})}\BibitemShut {NoStop}%
\bibitem [{\citenamefont {Wang}\ \emph {et~al.}(2015)\citenamefont {Wang},
  \citenamefont {Zhang}, \citenamefont {Corcovilos}, \citenamefont {Kumar},\
  and\ \citenamefont {Weiss}}]{Wang2015}%
  \BibitemOpen
  \bibfield  {author} {\bibinfo {author} {\bibfnamefont {Y.}~\bibnamefont
  {Wang}}, \bibinfo {author} {\bibfnamefont {X.}~\bibnamefont {Zhang}},
  \bibinfo {author} {\bibfnamefont {T.~A.}\ \bibnamefont {Corcovilos}},
  \bibinfo {author} {\bibfnamefont {A.}~\bibnamefont {Kumar}}, \ and\ \bibinfo
  {author} {\bibfnamefont {D.~S.}\ \bibnamefont {Weiss}},\ }\href {\doibase
  10.1103/PhysRevLett.115.043003} {\bibfield  {journal} {\bibinfo  {journal}
  {Phys. Rev. Lett.}\ }\textbf {\bibinfo {volume} {115}},\ \bibinfo {pages}
  {043003} (\bibinfo {year} {2015})}\BibitemShut {NoStop}%
\bibitem [{\citenamefont {Saffman}\ \emph {et~al.}(2010)\citenamefont
  {Saffman}, \citenamefont {Walker},\ and\ \citenamefont
  {M\o{}lmer}}]{Saffman2010}%
  \BibitemOpen
  \bibfield  {author} {\bibinfo {author} {\bibfnamefont {M.}~\bibnamefont
  {Saffman}}, \bibinfo {author} {\bibfnamefont {T.~G.}\ \bibnamefont {Walker}},
  \ and\ \bibinfo {author} {\bibfnamefont {K.}~\bibnamefont {M\o{}lmer}},\
  }\href {\doibase 10.1103/RevModPhys.82.2313} {\bibfield  {journal} {\bibinfo
  {journal} {Rev. Mod. Phys.}\ }\textbf {\bibinfo {volume} {82}},\ \bibinfo
  {pages} {2313} (\bibinfo {year} {2010})}\BibitemShut {NoStop}%
\bibitem [{\citenamefont {Weimer}\ \emph {et~al.}(2010)\citenamefont {Weimer},
  \citenamefont {Muller}, \citenamefont {Lesanovsky}, \citenamefont {Zoller},\
  and\ \citenamefont {B{\"{u}}chler}}]{Weimer2010}%
  \BibitemOpen
  \bibfield  {author} {\bibinfo {author} {\bibfnamefont {H.}~\bibnamefont
  {Weimer}}, \bibinfo {author} {\bibfnamefont {M.}~\bibnamefont {Muller}},
  \bibinfo {author} {\bibfnamefont {I.}~\bibnamefont {Lesanovsky}}, \bibinfo
  {author} {\bibfnamefont {P.}~\bibnamefont {Zoller}}, \ and\ \bibinfo {author}
  {\bibfnamefont {H.~P.}\ \bibnamefont {B{\"{u}}chler}},\ }\href
  {http://dx.doi.org/10.1038/nphys1614} {\bibfield  {journal} {\bibinfo
  {journal} {Nature Phys.}\ }\textbf {\bibinfo {volume} {6}},\ \bibinfo {pages}
  {382} (\bibinfo {year} {2010})}\BibitemShut {NoStop}%
\bibitem [{\citenamefont {Browaeys}\ \emph {et~al.}(2016)\citenamefont
  {Browaeys}, \citenamefont {Barredo},\ and\ \citenamefont
  {Lahaye}}]{Browaeys2016}%
  \BibitemOpen
  \bibfield  {author} {\bibinfo {author} {\bibfnamefont {A.}~\bibnamefont
  {Browaeys}}, \bibinfo {author} {\bibfnamefont {D.}~\bibnamefont {Barredo}}, \
  and\ \bibinfo {author} {\bibfnamefont {T.}~\bibnamefont {Lahaye}},\ }\href
  {\doibase 10.1088/0953-4075/49/15/152001} {\bibfield  {journal} {\bibinfo
  {journal} {J. Phys. B}\ }\textbf {\bibinfo {volume} {49}},\ \bibinfo {pages}
  {152001} (\bibinfo {year} {2016})}\BibitemShut {NoStop}%
\bibitem [{\citenamefont {Zeiher}\ \emph {et~al.}(2016)\citenamefont {Zeiher},
  \citenamefont {van Bijnen}, \citenamefont {Schauss}, \citenamefont {Hild},
  \citenamefont {Choi}, \citenamefont {Pohl}, \citenamefont {Bloch},\ and\
  \citenamefont {Gross}}]{Zeiher2016}%
  \BibitemOpen
  \bibfield  {author} {\bibinfo {author} {\bibfnamefont {J.}~\bibnamefont
  {Zeiher}}, \bibinfo {author} {\bibfnamefont {R.}~\bibnamefont {van Bijnen}},
  \bibinfo {author} {\bibfnamefont {P.}~\bibnamefont {Schauss}}, \bibinfo
  {author} {\bibfnamefont {S.}~\bibnamefont {Hild}}, \bibinfo {author}
  {\bibfnamefont {J.-Y.}\ \bibnamefont {Choi}}, \bibinfo {author}
  {\bibfnamefont {T.}~\bibnamefont {Pohl}}, \bibinfo {author} {\bibfnamefont
  {I.}~\bibnamefont {Bloch}}, \ and\ \bibinfo {author} {\bibfnamefont
  {C.}~\bibnamefont {Gross}},\ }\href {\doibase 10.1038/nphys3835} {\bibfield
  {journal} {\bibinfo  {journal} {Nature Phys.}\ }\textbf {\bibinfo {volume}
  {12}},\ \bibinfo {pages} {1095–} (\bibinfo {year} {2016})}\BibitemShut
  {NoStop}%
\bibitem [{\citenamefont {Jau}\ \emph {et~al.}(2016)\citenamefont {Jau},
  \citenamefont {Hankin}, \citenamefont {Keating}, \citenamefont {Deutsch},\
  and\ \citenamefont {Biedermann}}]{Biedermann2016}%
  \BibitemOpen
  \bibfield  {author} {\bibinfo {author} {\bibfnamefont {Y.-Y.}\ \bibnamefont
  {Jau}}, \bibinfo {author} {\bibfnamefont {A.~M.}\ \bibnamefont {Hankin}},
  \bibinfo {author} {\bibfnamefont {T.}~\bibnamefont {Keating}}, \bibinfo
  {author} {\bibfnamefont {I.~H.}\ \bibnamefont {Deutsch}}, \ and\ \bibinfo
  {author} {\bibfnamefont {G.~W.}\ \bibnamefont {Biedermann}},\ }\href
  {\doibase doi:10.1038/nphys3487} {\bibfield  {journal} {\bibinfo  {journal}
  {Nature Physics}\ }\textbf {\bibinfo {volume} {12}},\ \bibinfo {pages} {71}
  (\bibinfo {year} {2016})}\BibitemShut {NoStop}%
\bibitem [{\citenamefont {Barredo}\ \emph {et~al.}(2016)\citenamefont
  {Barredo}, \citenamefont {de~L\'es\'eleuc}, \citenamefont {Lienhard},
  \citenamefont {Lahaye},\ and\ \citenamefont {Browaeys}}]{Barredo2016}%
  \BibitemOpen
  \bibfield  {author} {\bibinfo {author} {\bibfnamefont {D.}~\bibnamefont
  {Barredo}}, \bibinfo {author} {\bibfnamefont {S.}~\bibnamefont
  {de~L\'es\'eleuc}}, \bibinfo {author} {\bibfnamefont {V.}~\bibnamefont
  {Lienhard}}, \bibinfo {author} {\bibfnamefont {T.}~\bibnamefont {Lahaye}}, \
  and\ \bibinfo {author} {\bibfnamefont {A.}~\bibnamefont {Browaeys}},\ }\href
  {\doibase 10.1126/science.aah3778} {\bibfield  {journal} {\bibinfo  {journal}
  {Science}\ }\textbf {\bibinfo {volume} {354}},\ \bibinfo {pages} {1021}
  (\bibinfo {year} {2016})}\BibitemShut {NoStop}%
\bibitem [{\citenamefont {Endres}\ \emph {et~al.}(2016)\citenamefont {Endres},
  \citenamefont {Bernien}, \citenamefont {Keesling}, \citenamefont {Levine},
  \citenamefont {Anschuetz}, \citenamefont {Krajenbrink}, \citenamefont
  {Senko}, \citenamefont {Vuletic}, \citenamefont {Greiner},\ and\
  \citenamefont {Lukin}}]{Endres2016}%
  \BibitemOpen
  \bibfield  {author} {\bibinfo {author} {\bibfnamefont {M.}~\bibnamefont
  {Endres}}, \bibinfo {author} {\bibfnamefont {H.}~\bibnamefont {Bernien}},
  \bibinfo {author} {\bibfnamefont {A.}~\bibnamefont {Keesling}}, \bibinfo
  {author} {\bibfnamefont {H.}~\bibnamefont {Levine}}, \bibinfo {author}
  {\bibfnamefont {E.~R.}\ \bibnamefont {Anschuetz}}, \bibinfo {author}
  {\bibfnamefont {A.}~\bibnamefont {Krajenbrink}}, \bibinfo {author}
  {\bibfnamefont {C.}~\bibnamefont {Senko}}, \bibinfo {author} {\bibfnamefont
  {V.}~\bibnamefont {Vuletic}}, \bibinfo {author} {\bibfnamefont
  {M.}~\bibnamefont {Greiner}}, \ and\ \bibinfo {author} {\bibfnamefont
  {M.~D.}\ \bibnamefont {Lukin}},\ }\href {\doibase 10.1126/science.aah3752}
  {\bibfield  {journal} {\bibinfo  {journal} {Science}\ }\textbf {\bibinfo
  {volume} {354}},\ \bibinfo {pages} {1024} (\bibinfo {year}
  {2016})}\BibitemShut {NoStop}%
\bibitem [{\citenamefont {Labuhn}\ \emph {et~al.}(2016)\citenamefont {Labuhn},
  \citenamefont {Barredo}, \citenamefont {Ravets}, \citenamefont
  {de~L\'es\'eleuc}, \citenamefont {Macr\`i}, \citenamefont {Lahaye},\ and\
  \citenamefont {Browaeys}}]{Labuhn2016}%
  \BibitemOpen
  \bibfield  {author} {\bibinfo {author} {\bibfnamefont {H.}~\bibnamefont
  {Labuhn}}, \bibinfo {author} {\bibfnamefont {D.}~\bibnamefont {Barredo}},
  \bibinfo {author} {\bibfnamefont {S.}~\bibnamefont {Ravets}}, \bibinfo
  {author} {\bibfnamefont {S.}~\bibnamefont {de~L\'es\'eleuc}}, \bibinfo
  {author} {\bibfnamefont {T.}~\bibnamefont {Macr\`i}}, \bibinfo {author}
  {\bibfnamefont {T.}~\bibnamefont {Lahaye}}, \ and\ \bibinfo {author}
  {\bibfnamefont {A.}~\bibnamefont {Browaeys}},\ }\href {\doibase
  10.1038/nature18274} {\bibfield  {journal} {\bibinfo  {journal} {Nature}\
  }\textbf {\bibinfo {volume} {534}},\ \bibinfo {pages} {667} (\bibinfo {year}
  {2016})}\BibitemShut {NoStop}%
\bibitem [{\citenamefont {Deng}\ \emph {et~al.}(2005)\citenamefont {Deng},
  \citenamefont {Porras},\ and\ \citenamefont {Cirac}}]{Deng2005}%
  \BibitemOpen
  \bibfield  {author} {\bibinfo {author} {\bibfnamefont {X.~L.}\ \bibnamefont
  {Deng}}, \bibinfo {author} {\bibfnamefont {D.}~\bibnamefont {Porras}}, \ and\
  \bibinfo {author} {\bibfnamefont {J.~I.}\ \bibnamefont {Cirac}},\ }\href
  {\doibase 10.1103/PhysRevA.72.063407} {\bibfield  {journal} {\bibinfo
  {journal} {Phys. Rev. A}\ }\textbf {\bibinfo {volume} {72}},\ \bibinfo
  {pages} {1} (\bibinfo {year} {2005})}\BibitemShut {NoStop}%
\bibitem [{\citenamefont {Hauke}\ \emph {et~al.}(2010)\citenamefont {Hauke},
  \citenamefont {Cucchietti}, \citenamefont {M{\"{u}}ller-Hermes},
  \citenamefont {Ba{\~{n}}uls}, \citenamefont {Cirac},\ and\ \citenamefont
  {Lewenstein}}]{Hauke2010}%
  \BibitemOpen
  \bibfield  {author} {\bibinfo {author} {\bibfnamefont {P.}~\bibnamefont
  {Hauke}}, \bibinfo {author} {\bibfnamefont {F.~M.}\ \bibnamefont
  {Cucchietti}}, \bibinfo {author} {\bibfnamefont {A.}~\bibnamefont
  {M{\"{u}}ller-Hermes}}, \bibinfo {author} {\bibfnamefont {M.~C.}\
  \bibnamefont {Ba{\~{n}}uls}}, \bibinfo {author} {\bibfnamefont {J.~I.}\
  \bibnamefont {Cirac}}, \ and\ \bibinfo {author} {\bibfnamefont
  {M.}~\bibnamefont {Lewenstein}},\ }\href {\doibase
  doi:10.1088/1367-2630/12/11/113037} {\bibfield  {journal} {\bibinfo
  {journal} {New J. Phys.}\ }\textbf {\bibinfo {volume} {12}},\ \bibinfo
  {pages} {113037} (\bibinfo {year} {2010})}\BibitemShut {NoStop}%
\bibitem [{\citenamefont {Varney}\ \emph {et~al.}(2011)\citenamefont {Varney},
  \citenamefont {Sun}, \citenamefont {Galitski},\ and\ \citenamefont
  {Rigol}}]{Varney2011}%
  \BibitemOpen
  \bibfield  {author} {\bibinfo {author} {\bibfnamefont {C.~N.}\ \bibnamefont
  {Varney}}, \bibinfo {author} {\bibfnamefont {K.}~\bibnamefont {Sun}},
  \bibinfo {author} {\bibfnamefont {V.}~\bibnamefont {Galitski}}, \ and\
  \bibinfo {author} {\bibfnamefont {M.}~\bibnamefont {Rigol}},\ }\href
  {\doibase 10.1103/PhysRevLett.107.077201} {\bibfield  {journal} {\bibinfo
  {journal} {Phys. Rev. Lett.}\ }\textbf {\bibinfo {volume} {107}},\ \bibinfo
  {pages} {077201} (\bibinfo {year} {2011})}\BibitemShut {NoStop}%
\bibitem [{\citenamefont {Peter}\ \emph {et~al.}(2012)\citenamefont {Peter},
  \citenamefont {M{\"{u}}ller}, \citenamefont {Wessel},\ and\ \citenamefont
  {B{\"{u}}chler}}]{Peter2012}%
  \BibitemOpen
  \bibfield  {author} {\bibinfo {author} {\bibfnamefont {D.}~\bibnamefont
  {Peter}}, \bibinfo {author} {\bibfnamefont {S.}~\bibnamefont {M{\"{u}}ller}},
  \bibinfo {author} {\bibfnamefont {S.}~\bibnamefont {Wessel}}, \ and\ \bibinfo
  {author} {\bibfnamefont {H.~P.}\ \bibnamefont {B{\"{u}}chler}},\ }\href
  {\doibase 10.1103/PhysRevLett.109.025303} {\bibfield  {journal} {\bibinfo
  {journal} {Phys. Rev. Lett.}\ }\textbf {\bibinfo {volume} {109}},\ \bibinfo
  {pages} {1} (\bibinfo {year} {2012})}\BibitemShut {NoStop}%
\bibitem [{\citenamefont {Yan}\ \emph {et~al.}(2013)\citenamefont {Yan},
  \citenamefont {Moses}, \citenamefont {Gadway}, \citenamefont {Covey},
  \citenamefont {Hazzard}, \citenamefont {Rey}, \citenamefont {Jin},\ and\
  \citenamefont {Ye}}]{Yan2013}%
  \BibitemOpen
  \bibfield  {author} {\bibinfo {author} {\bibfnamefont {B.}~\bibnamefont
  {Yan}}, \bibinfo {author} {\bibfnamefont {S.}~\bibnamefont {Moses}}, \bibinfo
  {author} {\bibfnamefont {B.}~\bibnamefont {Gadway}}, \bibinfo {author}
  {\bibfnamefont {J.}~\bibnamefont {Covey}}, \bibinfo {author} {\bibfnamefont
  {K.}~\bibnamefont {Hazzard}}, \bibinfo {author} {\bibfnamefont
  {A.}~\bibnamefont {Rey}}, \bibinfo {author} {\bibfnamefont {D.}~\bibnamefont
  {Jin}}, \ and\ \bibinfo {author} {\bibfnamefont {J.}~\bibnamefont {Ye}},\
  }\href {\doibase 10.1038/nature12483} {\bibfield  {journal} {\bibinfo
  {journal} {Nature}\ }\textbf {\bibinfo {volume} {501}},\ \bibinfo {pages}
  {521} (\bibinfo {year} {2013})}\BibitemShut {NoStop}%
\bibitem [{\citenamefont {Dalmonte}\ \emph {et~al.}(2015)\citenamefont
  {Dalmonte}, \citenamefont {Mirzaei}, \citenamefont {Muppalla}, \citenamefont
  {Marcos}, \citenamefont {Zoller},\ and\ \citenamefont
  {Kirchmair}}]{Dalmonte2015}%
  \BibitemOpen
  \bibfield  {author} {\bibinfo {author} {\bibfnamefont {M.}~\bibnamefont
  {Dalmonte}}, \bibinfo {author} {\bibfnamefont {S.~I.}\ \bibnamefont
  {Mirzaei}}, \bibinfo {author} {\bibfnamefont {P.~R.}\ \bibnamefont
  {Muppalla}}, \bibinfo {author} {\bibfnamefont {D.}~\bibnamefont {Marcos}},
  \bibinfo {author} {\bibfnamefont {P.}~\bibnamefont {Zoller}}, \ and\ \bibinfo
  {author} {\bibfnamefont {G.}~\bibnamefont {Kirchmair}},\ }\href {\doibase
  10.1103/PhysRevB.92.174507} {\bibfield  {journal} {\bibinfo  {journal} {Phys.
  Rev. B}\ }\textbf {\bibinfo {volume} {92}},\ \bibinfo {pages} {1} (\bibinfo
  {year} {2015})}\BibitemShut {NoStop}%
\bibitem [{\citenamefont {Mourachko}\ \emph {et~al.}(1998)\citenamefont
  {Mourachko}, \citenamefont {Comparat}, \citenamefont {de~Tomasi},
  \citenamefont {Fioretti}, \citenamefont {Nosbaum}, \citenamefont {Akulin},\
  and\ \citenamefont {Pillet}}]{Pillet1998}%
  \BibitemOpen
  \bibfield  {author} {\bibinfo {author} {\bibfnamefont {I.}~\bibnamefont
  {Mourachko}}, \bibinfo {author} {\bibfnamefont {D.}~\bibnamefont {Comparat}},
  \bibinfo {author} {\bibfnamefont {F.}~\bibnamefont {de~Tomasi}}, \bibinfo
  {author} {\bibfnamefont {A.}~\bibnamefont {Fioretti}}, \bibinfo {author}
  {\bibfnamefont {P.}~\bibnamefont {Nosbaum}}, \bibinfo {author} {\bibfnamefont
  {V.~M.}\ \bibnamefont {Akulin}}, \ and\ \bibinfo {author} {\bibfnamefont
  {P.}~\bibnamefont {Pillet}},\ }\href {\doibase 10.1103/PhysRevLett.80.253}
  {\bibfield  {journal} {\bibinfo  {journal} {Phys. Rev. Lett.}\ }\textbf
  {\bibinfo {volume} {80}},\ \bibinfo {pages} {253} (\bibinfo {year}
  {1998})}\BibitemShut {NoStop}%
\bibitem [{\citenamefont {Anderson}\ \emph {et~al.}(1998)\citenamefont
  {Anderson}, \citenamefont {Veale},\ and\ \citenamefont
  {Gallagher}}]{Gallagher1998}%
  \BibitemOpen
  \bibfield  {author} {\bibinfo {author} {\bibfnamefont {W.~R.}\ \bibnamefont
  {Anderson}}, \bibinfo {author} {\bibfnamefont {J.~R.}\ \bibnamefont {Veale}},
  \ and\ \bibinfo {author} {\bibfnamefont {T.~F.}\ \bibnamefont {Gallagher}},\
  }\href {\doibase 10.1103/PhysRevLett.80.249} {\bibfield  {journal} {\bibinfo
  {journal} {Phys. Rev. Lett.}\ }\textbf {\bibinfo {volume} {80}},\ \bibinfo
  {pages} {249} (\bibinfo {year} {1998})}\BibitemShut {NoStop}%
\bibitem [{\citenamefont {Barredo}\ \emph {et~al.}(2014)\citenamefont
  {Barredo}, \citenamefont {Ravets}, \citenamefont {Labuhn}, \citenamefont
  {B\'eguin}, \citenamefont {Vernier}, \citenamefont {Nogrette}, \citenamefont
  {Lahaye},\ and\ \citenamefont {Browaeys}}]{Barredo2014}%
  \BibitemOpen
  \bibfield  {author} {\bibinfo {author} {\bibfnamefont {D.}~\bibnamefont
  {Barredo}}, \bibinfo {author} {\bibfnamefont {S.}~\bibnamefont {Ravets}},
  \bibinfo {author} {\bibfnamefont {H.}~\bibnamefont {Labuhn}}, \bibinfo
  {author} {\bibfnamefont {L.}~\bibnamefont {B\'eguin}}, \bibinfo {author}
  {\bibfnamefont {A.}~\bibnamefont {Vernier}}, \bibinfo {author} {\bibfnamefont
  {F.}~\bibnamefont {Nogrette}}, \bibinfo {author} {\bibfnamefont
  {T.}~\bibnamefont {Lahaye}}, \ and\ \bibinfo {author} {\bibfnamefont
  {A.}~\bibnamefont {Browaeys}},\ }\href {\doibase
  10.1103/PhysRevLett.112.183002} {\bibfield  {journal} {\bibinfo  {journal}
  {Phys. Rev. Lett.}\ }\textbf {\bibinfo {volume} {112}},\ \bibinfo {pages}
  {183002} (\bibinfo {year} {2014})}\BibitemShut {NoStop}%
\bibitem [{\citenamefont {Maxwell}\ \emph {et~al.}(2014)\citenamefont
  {Maxwell}, \citenamefont {Szwer}, \citenamefont {Paredes-Barato},
  \citenamefont {Busche}, \citenamefont {Pritchard}, \citenamefont {Gauguet},
  \citenamefont {Jones},\ and\ \citenamefont {Adams}}]{Maxwell2014}%
  \BibitemOpen
  \bibfield  {author} {\bibinfo {author} {\bibfnamefont {D.}~\bibnamefont
  {Maxwell}}, \bibinfo {author} {\bibfnamefont {D.~J.}\ \bibnamefont {Szwer}},
  \bibinfo {author} {\bibfnamefont {D.}~\bibnamefont {Paredes-Barato}},
  \bibinfo {author} {\bibfnamefont {H.}~\bibnamefont {Busche}}, \bibinfo
  {author} {\bibfnamefont {J.~D.}\ \bibnamefont {Pritchard}}, \bibinfo {author}
  {\bibfnamefont {A.}~\bibnamefont {Gauguet}}, \bibinfo {author} {\bibfnamefont
  {M.~P.~A.}\ \bibnamefont {Jones}}, \ and\ \bibinfo {author} {\bibfnamefont
  {C.~S.}\ \bibnamefont {Adams}},\ }\href {\doibase 10.1103/PhysRevA.89.043827}
  {\bibfield  {journal} {\bibinfo  {journal} {Phys. Rev. A}\ }\textbf {\bibinfo
  {volume} {89}},\ \bibinfo {pages} {1} (\bibinfo {year} {2014})}\BibitemShut
  {NoStop}%
\bibitem [{\citenamefont {Orioli}\ \emph {et~al.}(2017)\citenamefont {Orioli},
  \citenamefont {Signoles}, \citenamefont {Wildhagen}, \citenamefont
  {G{\"{u}}nter}, \citenamefont {Berges}, \citenamefont {Whitlock},\ and\
  \citenamefont {Weidem{\"{u}}ller}}]{Orioli2017}%
  \BibitemOpen
  \bibfield  {author} {\bibinfo {author} {\bibfnamefont {A.~P.}\ \bibnamefont
  {Orioli}}, \bibinfo {author} {\bibfnamefont {A.}~\bibnamefont {Signoles}},
  \bibinfo {author} {\bibfnamefont {H.}~\bibnamefont {Wildhagen}}, \bibinfo
  {author} {\bibfnamefont {G.}~\bibnamefont {G{\"{u}}nter}}, \bibinfo {author}
  {\bibfnamefont {J.}~\bibnamefont {Berges}}, \bibinfo {author} {\bibfnamefont
  {S.}~\bibnamefont {Whitlock}}, \ and\ \bibinfo {author} {\bibfnamefont
  {M.}~\bibnamefont {Weidem{\"{u}}ller}},\ }\href
  {http://arxiv.org/abs/1703.05957} {\bibfield  {journal} {\bibinfo  {journal}
  {arXiv:1703.05957}\ } (\bibinfo {year} {2017})}\BibitemShut {NoStop}%
\bibitem [{\citenamefont {B\'eguin}\ \emph {et~al.}(2013)\citenamefont
  {B\'eguin}, \citenamefont {Vernier}, \citenamefont {Chicireanu},
  \citenamefont {Lahaye},\ and\ \citenamefont {Browaeys}}]{Beguin2013}%
  \BibitemOpen
  \bibfield  {author} {\bibinfo {author} {\bibfnamefont {L.}~\bibnamefont
  {B\'eguin}}, \bibinfo {author} {\bibfnamefont {A.}~\bibnamefont {Vernier}},
  \bibinfo {author} {\bibfnamefont {R.}~\bibnamefont {Chicireanu}}, \bibinfo
  {author} {\bibfnamefont {T.}~\bibnamefont {Lahaye}}, \ and\ \bibinfo {author}
  {\bibfnamefont {A.}~\bibnamefont {Browaeys}},\ }\href {\doibase
  10.1103/PhysRevLett.110.263201} {\bibfield  {journal} {\bibinfo  {journal}
  {Phys. Rev. Lett.}\ }\textbf {\bibinfo {volume} {110}},\ \bibinfo {pages}
  {263201} (\bibinfo {year} {2013})}\BibitemShut {NoStop}%
\bibitem [{\citenamefont {Nogrette}\ \emph {et~al.}(2014)\citenamefont
  {Nogrette}, \citenamefont {Labuhn}, \citenamefont {Ravets}, \citenamefont
  {Barredo}, \citenamefont {B\'eguin}, \citenamefont {Vernier}, \citenamefont
  {Lahaye},\ and\ \citenamefont {Browaeys}}]{Nogrette2014}%
  \BibitemOpen
  \bibfield  {author} {\bibinfo {author} {\bibfnamefont {F.}~\bibnamefont
  {Nogrette}}, \bibinfo {author} {\bibfnamefont {H.}~\bibnamefont {Labuhn}},
  \bibinfo {author} {\bibfnamefont {S.}~\bibnamefont {Ravets}}, \bibinfo
  {author} {\bibfnamefont {D.}~\bibnamefont {Barredo}}, \bibinfo {author}
  {\bibfnamefont {L.}~\bibnamefont {B\'eguin}}, \bibinfo {author}
  {\bibfnamefont {A.}~\bibnamefont {Vernier}}, \bibinfo {author} {\bibfnamefont
  {T.}~\bibnamefont {Lahaye}}, \ and\ \bibinfo {author} {\bibfnamefont
  {A.}~\bibnamefont {Browaeys}},\ }\href {\doibase 10.1103/PhysRevX.4.021034}
  {\bibfield  {journal} {\bibinfo  {journal} {Phys. Rev. X}\ }\textbf {\bibinfo
  {volume} {4}},\ \bibinfo {pages} {021034} (\bibinfo {year}
  {2014})}\BibitemShut {NoStop}%
\bibitem [{\citenamefont {{\v{S}}ibali{\'{c}}}\ \emph
  {et~al.}(2016)\citenamefont {{\v{S}}ibali{\'{c}}}, \citenamefont {Pritchard},
  \citenamefont {Adams},\ and\ \citenamefont {Weatherill}}]{Sibalic2016}%
  \BibitemOpen
  \bibfield  {author} {\bibinfo {author} {\bibfnamefont {N.}~\bibnamefont
  {{\v{S}}ibali{\'{c}}}}, \bibinfo {author} {\bibfnamefont {J.~D.}\
  \bibnamefont {Pritchard}}, \bibinfo {author} {\bibfnamefont {C.~S.}\
  \bibnamefont {Adams}}, \ and\ \bibinfo {author} {\bibfnamefont {K.~J.}\
  \bibnamefont {Weatherill}},\ }\href {http://arxiv.org/abs/1612.05529}
  {\bibfield  {journal} {\bibinfo  {journal} {arXiv:1612.05529}\ } (\bibinfo
  {year} {2016})}\BibitemShut {NoStop}%
\bibitem [{\citenamefont {Weber}\ \emph {et~al.}(2016)\citenamefont {Weber},
  \citenamefont {Tresp}, \citenamefont {Menke}, \citenamefont {Urvoy},
  \citenamefont {Firstenberg}, \citenamefont {B{\"{u}}chler},\ and\
  \citenamefont {Hofferberth}}]{Weber2016}%
  \BibitemOpen
  \bibfield  {author} {\bibinfo {author} {\bibfnamefont {S.}~\bibnamefont
  {Weber}}, \bibinfo {author} {\bibfnamefont {C.}~\bibnamefont {Tresp}},
  \bibinfo {author} {\bibfnamefont {H.}~\bibnamefont {Menke}}, \bibinfo
  {author} {\bibfnamefont {A.}~\bibnamefont {Urvoy}}, \bibinfo {author}
  {\bibfnamefont {O.}~\bibnamefont {Firstenberg}}, \bibinfo {author}
  {\bibfnamefont {H.~P.}\ \bibnamefont {B{\"{u}}chler}}, \ and\ \bibinfo
  {author} {\bibfnamefont {S.}~\bibnamefont {Hofferberth}},\ }\href
  {https://arxiv.org/pdf/1612.08053.pdf} {\bibfield  {journal} {\bibinfo
  {journal} {arXiv:1612.08053}\ } (\bibinfo {year} {2016})}\BibitemShut
  {NoStop}%
\bibitem [{\citenamefont {Saffman}\ and\ \citenamefont
  {Walker}(2005)}]{Saffman2005}%
  \BibitemOpen
  \bibfield  {author} {\bibinfo {author} {\bibfnamefont {M.}~\bibnamefont
  {Saffman}}\ and\ \bibinfo {author} {\bibfnamefont {T.~G.}\ \bibnamefont
  {Walker}},\ }\href {\doibase 10.1103/PhysRevA.72.022347} {\bibfield
  {journal} {\bibinfo  {journal} {Phys. Rev. A}\ }\textbf {\bibinfo {volume}
  {72}},\ \bibinfo {pages} {022347} (\bibinfo {year} {2005})}\BibitemShut
  {NoStop}%
\bibitem [{\citenamefont {Li}\ \emph {et~al.}(2013)\citenamefont {Li},
  \citenamefont {Dudin},\ and\ \citenamefont {Kuzmich}}]{Kuzmich2013}%
  \BibitemOpen
  \bibfield  {author} {\bibinfo {author} {\bibfnamefont {L.}~\bibnamefont
  {Li}}, \bibinfo {author} {\bibfnamefont {Y.~O.}\ \bibnamefont {Dudin}}, \
  and\ \bibinfo {author} {\bibfnamefont {A.}~\bibnamefont {Kuzmich}},\ }\href
  {\doibase 10.1038/nature12227} {\bibfield  {journal} {\bibinfo  {journal}
  {Nature}\ }\textbf {\bibinfo {volume} {498}},\ \bibinfo {pages} {466}
  (\bibinfo {year} {2013})}\BibitemShut {NoStop}%
\bibitem [{Note1()}]{Note1}%
  \BibitemOpen
  \bibinfo {note} {The beam waist was determined by using the addressing beam
  as a single-atom trap. The trap depth is measured by Rydberg spectroscopy
  \cite {Labuhn2014} and the trapping frequency by parametric heating \cite
  {Piotrowicz2013}.}\BibitemShut {Stop}%
\bibitem [{Note2()}]{Note2}%
  \BibitemOpen
  \bibinfo {note} {High Finesse, WLM SU10}\BibitemShut {NoStop}%
\bibitem [{\citenamefont {Younge}\ \emph {et~al.}(2010)\citenamefont {Younge},
  \citenamefont {Knuffman}, \citenamefont {Anderson},\ and\ \citenamefont
  {Raithel}}]{Raithel2010}%
  \BibitemOpen
  \bibfield  {author} {\bibinfo {author} {\bibfnamefont {K.~C.}\ \bibnamefont
  {Younge}}, \bibinfo {author} {\bibfnamefont {B.}~\bibnamefont {Knuffman}},
  \bibinfo {author} {\bibfnamefont {S.~E.}\ \bibnamefont {Anderson}}, \ and\
  \bibinfo {author} {\bibfnamefont {G.}~\bibnamefont {Raithel}},\ }\href
  {\doibase 10.1103/PhysRevLett.104.173001} {\bibfield  {journal} {\bibinfo
  {journal} {Phys. Rev. Lett.}\ }\textbf {\bibinfo {volume} {104}},\ \bibinfo
  {pages} {173001} (\bibinfo {year} {2010})}\BibitemShut {NoStop}%
\bibitem [{\citenamefont {Gomez}\ \emph {et~al.}(2004)\citenamefont {Gomez},
  \citenamefont {Aubin}, \citenamefont {Orozco},\ and\ \citenamefont
  {Sprouse}}]{Gomez2004}%
  \BibitemOpen
  \bibfield  {author} {\bibinfo {author} {\bibfnamefont {E.}~\bibnamefont
  {Gomez}}, \bibinfo {author} {\bibfnamefont {S.}~\bibnamefont {Aubin}},
  \bibinfo {author} {\bibfnamefont {L.~A.}\ \bibnamefont {Orozco}}, \ and\
  \bibinfo {author} {\bibfnamefont {G.~D.}\ \bibnamefont {Sprouse}},\ }\href
  {\doibase 10.1364/JOSAB.21.002058} {\bibfield  {journal} {\bibinfo  {journal}
  {J. Opt. Soc. Am. B}\ }\textbf {\bibinfo {volume} {21}},\ \bibinfo {pages}
  {2058} (\bibinfo {year} {2004})}\BibitemShut {NoStop}%
\bibitem [{\citenamefont {Wilk}\ \emph {et~al.}(2010)\citenamefont {Wilk},
  \citenamefont {Ga\"etan}, \citenamefont {Evellin}, \citenamefont {Wolters},
  \citenamefont {Miroshnychenko}, \citenamefont {Grangier},\ and\ \citenamefont
  {Browaeys}}]{Wilk2010}%
  \BibitemOpen
  \bibfield  {author} {\bibinfo {author} {\bibfnamefont {T.}~\bibnamefont
  {Wilk}}, \bibinfo {author} {\bibfnamefont {A.}~\bibnamefont {Ga\"etan}},
  \bibinfo {author} {\bibfnamefont {C.}~\bibnamefont {Evellin}}, \bibinfo
  {author} {\bibfnamefont {J.}~\bibnamefont {Wolters}}, \bibinfo {author}
  {\bibfnamefont {Y.}~\bibnamefont {Miroshnychenko}}, \bibinfo {author}
  {\bibfnamefont {P.}~\bibnamefont {Grangier}}, \ and\ \bibinfo {author}
  {\bibfnamefont {A.}~\bibnamefont {Browaeys}},\ }\href {\doibase
  10.1103/PhysRevLett.104.010502} {\bibfield  {journal} {\bibinfo  {journal}
  {Phys. Rev. Lett.}\ }\textbf {\bibinfo {volume} {104}},\ \bibinfo {pages}
  {010502} (\bibinfo {year} {2010})}\BibitemShut {NoStop}%
\bibitem [{\citenamefont {Viscor}(2017)}]{Viscor2017}%
  \BibitemOpen
  \bibfield  {author} {\bibinfo {author} {\bibfnamefont {D.}~\bibnamefont
  {Viscor}},\ }\href@noop {} {\bibfield  {journal} {\bibinfo  {journal} {in
  preparation}\ } (\bibinfo {year} {2017})}\BibitemShut {NoStop}%
\bibitem [{\citenamefont {Glaetzle}\ \emph {et~al.}(2017)\citenamefont
  {Glaetzle}, \citenamefont {Ender}, \citenamefont {Wild}, \citenamefont
  {Choi}, \citenamefont {Pichler}, \citenamefont {Lukin},\ and\ \citenamefont
  {Zoller}}]{Zoller2017}%
  \BibitemOpen
  \bibfield  {author} {\bibinfo {author} {\bibfnamefont {A.~W.}\ \bibnamefont
  {Glaetzle}}, \bibinfo {author} {\bibfnamefont {K.}~\bibnamefont {Ender}},
  \bibinfo {author} {\bibfnamefont {D.}~\bibnamefont {Wild}}, \bibinfo {author}
  {\bibfnamefont {S.}~\bibnamefont {Choi}}, \bibinfo {author} {\bibfnamefont
  {H.}~\bibnamefont {Pichler}}, \bibinfo {author} {\bibfnamefont
  {M.}~\bibnamefont {Lukin}}, \ and\ \bibinfo {author} {\bibfnamefont
  {P.}~\bibnamefont {Zoller}},\ }\href {https://arxiv.org/pdf/1704.08837.pdf}
  {\bibfield  {journal} {\bibinfo  {journal} {arXiv:1704.08837}\ } (\bibinfo
  {year} {2017})}\BibitemShut {NoStop}%
\bibitem [{\citenamefont {Piotrowicz}\ \emph {et~al.}(2013)\citenamefont
  {Piotrowicz}, \citenamefont {Lichtman}, \citenamefont {Maller}, \citenamefont
  {Li}, \citenamefont {Zhang}, \citenamefont {Isenhower},\ and\ \citenamefont
  {Saffman}}]{Piotrowicz2013}%
  \BibitemOpen
  \bibfield  {author} {\bibinfo {author} {\bibfnamefont {M.~J.}\ \bibnamefont
  {Piotrowicz}}, \bibinfo {author} {\bibfnamefont {M.}~\bibnamefont
  {Lichtman}}, \bibinfo {author} {\bibfnamefont {K.}~\bibnamefont {Maller}},
  \bibinfo {author} {\bibfnamefont {G.}~\bibnamefont {Li}}, \bibinfo {author}
  {\bibfnamefont {S.}~\bibnamefont {Zhang}}, \bibinfo {author} {\bibfnamefont
  {L.}~\bibnamefont {Isenhower}}, \ and\ \bibinfo {author} {\bibfnamefont
  {M.}~\bibnamefont {Saffman}},\ }\href {\doibase 10.1103/PhysRevA.88.013420}
  {\bibfield  {journal} {\bibinfo  {journal} {Phys. Rev. A}\ }\textbf {\bibinfo
  {volume} {88}},\ \bibinfo {pages} {013420} (\bibinfo {year}
  {2013})}\BibitemShut {NoStop}%
\end{thebibliography}%

\end{document}